\documentclass[11pt]{article}
\begin{document}
\title
{Trace of phase-space noncommutativity in the response of a free particle to linearized gravitational waves}
\author{
{\bf {\normalsize Sunandan Gangopadhyay}$^{a,b}
$\thanks{sunandan.gangopadhyay@gmail.com}},
{\bf {\normalsize Anirban Saha}
$^{a,b}$\thanks{anirban@iucaa.ernet.in}},
{\bf {\normalsize Swarup Saha}$^{a}$\thanks{saha18swarup@gmail.com}}\\
$^{a}$ {\normalsize Department of Physics, West Bengal State University, Barasat, India}\\
$^{b}${\normalsize Visiting Associate in Inter University Centre for Astronomy $\&$ Astrophysics,}\\
{\normalsize Pune, India}\\[0.3cm]
}
\date{}

\maketitle

\begin{abstract}
\noindent Interaction of linearized gravitational waves with a otherwise free particle has been studied quantum mechanically in a noncommutative phase-space to examine whether the particle's response to the gravitational wave gets modified due to spatial and/or momentum noncommutativity. The result shows that momentum noncommutativity introduces a oscillatory noise with a specific frequency determined by the fundamental momentum scale and particle mass. Because of the global nature of the phase-space noncommutativity such noise will have similar characteristics for all detector sites and thus will stand out in a data cross-correlation procedure. If detected, this noise will provide evidence of momentum noncommutativity and also an estimation of the relevant noncommutative parameter. 
\end{abstract}

\maketitle
%%%%%%%%%%%%%%%%%%%%%%%%%%%%%%%%%%%%%%%%%%%%%%%%%%%%%%%%%%%%%%
At the Planck scale the space-time is thought to have a granular structure \cite{Dop1, Dop2, Alu} much like the phase-space of quantum mechanics (QM). This granularity can be theoretically realized by describing the space-time with a set of coordinates $x^{\mu}, \left( \mu = 0,1,2,3 \right)$ following a noncommutative (NC) algebra $\left[x^{\mu}, x^{\nu}\right] = i \theta^{\mu \nu}$, where $\theta^{\mu \nu}$ is a constant symmetric tensor of second rank. This space-time is referred to as canonical NC space-time \cite{snyder}. Among other types of noncommutativity most notable are associated with Quantum group theory \cite{majid}, e.g., Lie-algebric $\kappa$-Minkowski noncommutativity \cite{ref1, ref2, ref3}. Theories defined on such NC spaces are generically called NC theories \cite{sw, szabo, ref4}. In recent years, there have been speculations of extending the NC space to a more general NC phase-space\footnote{Since we only consider the NC quantum mechanical phase-space we choose to ignore noncommutativity among the spatial coordinates and time.} \cite{momentum_NC1, momentum_NC2, momentum_NC3, samanta} with both the canonical pairs following the algebra: 
\begin{eqnarray}
\left[{\hat x}_{i}, {\hat p}_{j}\right] = i\tilde{h}\delta_{ij} \>, \quad 
\left[{\hat x}_{i}, {\hat x}_{j}\right] = i \theta \epsilon_{ij} \>,\quad 
\left[{\hat p}_{i}, {\hat p}_{j}\right] =i \bar{\theta} \epsilon_{ij} \>
\label{e9a}
\end{eqnarray}
where $i,j = 1, 2; \tilde{h}= \hbar\left({1} + \frac{\theta\bar{\theta}}{4\hbar^2}\right)$ is the modified Planck's constant; $\theta$ and $\bar{\theta}$ are spatial and momentum NC parameters respectively and $\epsilon_{ij}= -\epsilon_{ji}, (\epsilon_{12}=1)$ is the anti-symmetric tensor in two dimensions\footnote{For our purpose we can confine our attention to the four-dimensional NC phase-space}. The main argument for considering a NC phase-space is that noncommutativity between momenta arises naturally as a consequence of noncommutativity between coordinates, as momenta are defined as the partial derivatives of the action with respect to the coordinates \cite{momentum_NC4}. Theories defined over a noncommutative phase-space has also been furnished in \cite{giri, bensg, dulat} in context of the NC harmonic oscillator and NC Lorentz transformations. 

Apart from the mathematical complexities involved, the biggest challenge in any NC theory is to identify experimentally detectable effects of noncommutativity owing to the extreme smallness of the NC parameters $\theta $ and $\bar{\theta} $ appearing in the algebra (\ref{e9a}). Although effects of the NC structure of space-time may appear near the string/Planckian scale, it is hoped that some low energy relics of such effects may exist and their phenomenological consequences are currently being explored at the level of quantum mechanics \cite{Nair, bcsgas, pmh, bert0, ani, galileo, mezin, hazra, gov}. 
%Signatures of noncommutativity and/or the bounds on the NC scale come from neutrino astrophysics \cite{nc_astro} and cosmology \cite{nc_cosmo}; and also from high energy particle physics \cite{nc_hep}, producing a scale of noncommutativity of order few TeV's. 
Typical low energy non-accelerator experiments are the Lamb-shift \cite{cst}, and clock-comparison experiments \cite{carol} where the upper bound on the value of the canonical NC parameter was found to be $\theta\leq \left(10 {\rm TeV}\right)^{-2}$ which corresponds to $4 \times 10^{-40} {\rm m}^{2}$ for $\hbar$$=$$c$$=$$1$. In theories with lie-algebra valued NC geometry the NC parameter is recently estimated to be much higher in the energy scale, ranging in $10^{22} - 10^{24}$Tev \cite{ref3}. Whereas such upper bounds on the momentum NC parameter \cite{bert0, samanta} is $\bar{\theta} \leq 2.32 \times 10^{-61} {\rm kg^{2}m^{2}sec^{-2}}$ and time-space NC parameter \cite{ani} is $\theta^{0i} \leq 9.51\times 10^{-18} {\rm m}^{2}$ which are shown to be mutually consistent in \cite{ani}. Also, recent studies in NC quantum mechanics revealed that the NC parameter associated with different particles are not same \cite{pmh} and this bound could be as high as $\theta \leq \left(4 {\rm GeV}\right)^{-2} - \left(30 {\rm MeV}\right)^{-2}$ \cite{stern}. These upper bounds correspond to the length scale range $\sim 10^{-20} {\rm m} - 10^{-17} {\rm m}$ and thus suggest the potential possibility of finding the NC signature in present day high-precision gravitational wave detection experiments.

With the development of various ground based GW detectors\cite{Magg} like LIGO \cite{abramovici}, VIRGO \cite{caron}, GEO \cite{luck} and TAMA \cite{ando}, the possibility of direct detection of GW(s) with a strain sensitivity of the order of $\frac{\delta L}{L} \approx h \sim 10^{-21}/\sqrt{{\rm{Hz}}} $ or better in the frequency range between 100-1000 Hz is expected in the near future \cite{thorn}, where $L \sim 1{\rm km}$ is the GW interferometer cavity arm length and $\delta L \sim 10^{-18} {\rm m}$ is its variation due to the passing GW. This roughly means that these modern GW interferometers are capable of monitoring the (relative) positions of their test masses (the beam splitter and the mirrors) with an accuracy of order $10^{-18} $m and better \cite{ligoprototype}. Also, GW detectors based on matter-wave interferometry \cite{atom_int, atom_int1, atom_int2} have been suggested recently with similar strain sensitivity. These interferometers are thus ideally suited to monitor the fuzziness introduced in distance measurements between test masses due to NC nature of phase-space. A NC theory where the concept of distance is fundamentally fuzzy, even in the idealized situation where all classical and ordinary quantum mechanical noise sources are completely eliminated, the read-out of an interferometer would still be noisy as a result of phase-space noncommutativity \cite{nature, cam}. 

Since GW affects matter at very small length-scale \cite{Caves} and NC phase-space structure is inherently quantum mechanical in nature, a quantum mechanical theory of the GW-matter interaction in an NC phase-space would be necessary to predict the possible NC noise sources in the GW detector read-outs. With this motivation we have recently adopted a systematic approach \cite{speli} to study the effect of GW(s) in the long-wavelength and low velocity limit on the test matter, e.g., free particle and a harmonic oscillator \cite{ncgw1, ncgw2}, in a noncommutative (NC) space. 

In the present paper, we extend our earlier study \cite{ncgw1} of a free test particle under linearized GW(s) by considering a NC phase-space, where both coordinate and momentum are assumed to follow NC algebra (\ref{e9a}) instead of a NC space where only coordinates are noncommutative. The motivation of this generalization to NC phase-space is the following: In \cite{ncgw1} we have shown that the NC signature caused by spatial noncommutativity in this system originates from the matter-GW interaction term and so comes coupled with the GW amplitude. Therefore, owing to the small amplitude of GW and even smaller upper-bounds of the spatial NC parameter, the spatial NC effects may be well below the detection range of the present GW detectors. But, in the present paper, apart from spatial NC sector, we additionally consider NC effect caused by the momentum noncommutativity which arises not only from the interaction term but also from the kinetic term. So the present system will have pure NC signature of momentum noncommutativity, linear in only the momentum NC parameter and independent of any gravitational coupling and thus may present a better chance of detection. 

 To begin with, we consider a NC phase-space algebra (\ref{e9a}) that can be mapped into the commutative Heisenberg-Weyl algebra between variables $(X,P)$ through the following linear map from $(\hat x, \hat p)$ to $(X,Y)$
\begin{eqnarray}
{\hat x}_{i} = X_{i} - \frac{1}{2 \hbar} 
\theta \epsilon_{ij} P_{j}\>, \quad {\hat p}_{i} = P_{i} + \frac{1}{2 \hbar} 
\bar{\theta} \epsilon_{ij} X_{j}\>\>.
\label{e9b}
\end{eqnarray}
This map enables us to describe the NC system in terms of an effective commutative-equivalent model where the variables $(X,P)$ follow the ordinary Heisenberg-Weyl algebra and NC effect is manifest as a perturbative coupling. With this background, we now construct the quantum mechanics of a test particle interacting with linearized GW(s) in NC phase-space. In our earlier work \cite{ncgw1, ncgw2} we have presented the argument of working back from the geodesic deviation equation in the proper detector frame to the corresponding Hamiltonian. Following the same path, the NC Hamiltonian in the present case becomes
\begin{equation}
\hat{H} = \frac{1}{2m}{\hat p}_{j}^{2} +  \Gamma^j_{0k} {\hat x}_{j}{\hat p}_{k}
\label{e9}
\end{equation}
where the notations have their usual meaning. Using eq.(\ref{e9b}) and the traceless property of the GW, we rewrite the above Hamiltonian in terms of the operators $(X,P)$
\begin{eqnarray}
\hat{H} = \frac{ P_{j}{}^{2}}{2m} + \Gamma^j_{0k} X_{j} P_{k} + \frac{\bar{\theta}}{2 m\hbar} \epsilon_{jm} X^{j} P_{m}  
- \frac{\theta }{2 \hbar} \epsilon_{jm}  P_{m} P_{k}  \Gamma^j_{0k} \>
+\frac{{\bar{\theta}} }{2 \hbar} \epsilon_{jm} X_{m} X_{k}  \Gamma^j_{0k}\>. 
\label{e12}
\end{eqnarray}
This is the Hamiltonian of the system in the NC phase-space. Since it has been demonstrated in various formulations of NC general relativity \cite{grav, banerjee11} that any NC correction in the gravity sector is second order in the NC parameter, therefore, in a first order theory in NC phase-space, the linearised GW remains unaltered by NC effects and any NC correction appearing in the system will be through the particle sector only. 
We re-express the NC Hamiltonian in eq.(\ref{e12}) in terms of the raising and lowering operators defined by 
\begin{eqnarray}
X_j = \left({\hbar\over 2m\varpi}\right)^{1/2}
\left(a_j+a_j^\dagger\right)\quad;\quad
P_j = -i\left({\hbar m\varpi\over 2}\right)^{1/2} 
\left(a_j-a_j^\dagger\right)
\label{xp}
\end{eqnarray}
with the frequency $\varpi$ determined from the initial uncertainty in either the position or the momentum of the particle \cite{speli}.
This gives  
\begin{eqnarray}
\hat{H} &=& \frac {\hbar\varpi}{4}\left(2 a_j^\dagger a_j + 1 -a_j^2 - a_j^{\dagger 2}\right) - \frac{i\hbar}{4} \dot h_{jk}\left(a_j a_k - a_j^\dagger a_k^\dagger\right)+ \frac {i\bar{\theta}}{4m}\epsilon_{jk}a_j^\dagger a_k  \nonumber\\
&&+\frac{m\varpi\theta}{8}\epsilon_{jm}\dot{h}_{jk}(a_m a_k -a_m a_{k}^{\dagger} +C.C.) 
+\frac{\bar\theta}{8m\varpi}\epsilon_{jm}\dot{h}_{jk}(a_m a_k +a_m a_{k}^{\dagger} +C.C.).
\label{e16}
\end{eqnarray}
which is used to compute the time evolution of $a_{j}(t)$  
\begin{eqnarray}
\frac{da_{j}(t)}{dt} &=& \frac {-i\varpi}{2}\left(a_j-a_j^\dagger\right) + 
\frac{1}{2}\dot h_{jk}a^\dagger_k + 
\frac{\bar{\theta}}{4m\hbar}\epsilon_{jk} a_{k}+\frac{im\varpi\theta}{8\hbar}
(\epsilon_{lj}\dot h_{lk}+\epsilon_{lk}\dot h_{lj})(a_k - a_{k}^{\dagger})\nonumber\\
&&-\frac{i\bar{\theta}}{8m\varpi\hbar}
(\epsilon_{lj}\dot h_{lk}+\epsilon_{lk}\dot h_{lj})(a_k + a_{k}^{\dagger})
\label{e17}
\end{eqnarray}
and that of $a_{j}^{\dagger}(t)$, given by the C.C of the above equation.
Note that the raising and lowering operators satisfy the commutation relations
\begin{eqnarray}
\left[a_j(t), a^\dagger_k(t)\right] = \delta_{jk}\quad;\quad
\left[a_j(t), a_k(t)\right] = 0 = \left[a^\dagger_j(t), a^\dagger_k(t)\right].
\label{e18}
\end{eqnarray}
We now make use of the time dependent Bogoliubov transformations which relate $a_j(t)$ 
and $a_j^{\dagger}(t)$ in terms of the operators at time $t=0$
\begin{eqnarray}
a_j(t) = u_{jk}(t) a_k(0) + v_{jk}(t)a^\dagger_k(0)\quad;\quad 
a_j^\dagger(t) = a_k^\dagger(0)\bar u_{kj}(t)  + a_k(0)\bar v_{kj}(t)\>
\label{e19}
\end{eqnarray}
where  $u_{jk}$ and $v_{jk}$ are the generalized Bogoliubov coefficients. Since $a_j(t = 0) = a_j(0)$, $u_{jk}(t)$ and $v_{jk}(t)$ have the boundary conditions 
\begin{eqnarray}
u_{jk}(0) =  I  \quad;\quad  v_{jk}(0) = 0~.
\label{bc0}
\end{eqnarray}
From eq(s)(\ref{e17}, \ref{e19}), we get the following equations of motion 
in terms of $\zeta = u - v^\dagger$ and $\xi = u + v^\dagger$:
\begin{eqnarray}
\frac{d\zeta_{jk}}{dt}=
-\frac{1}{2}{\dot h}_{jl}\zeta_{lk} + \frac {\bar{\theta}}{4m\hbar}\epsilon_{jl}\zeta_{lk}
- \frac{i\bar{\theta}}{4m\varpi\hbar}
(\epsilon_{lj}\dot h_{lp}+\epsilon_{lp}\dot h_{lj})\xi_{pk}\>
\label{e21a}\\
\frac{d \xi_{jk}}{dt} = -i\varpi \zeta_{jk} + 
\frac{1}{2}{\dot h}_{jl}\xi_{lk} 
+\frac{im\varpi\theta}{4\hbar}(\epsilon_{pl}\dot h_{jp}-\epsilon_{jp}\dot h_{pl})\zeta_{lk}
+\frac{\bar{\theta}}{4m\hbar}\epsilon_{jl}\xi_{lk}~.
\label{e21b} 
\end{eqnarray}
We shall now solve the above equation for the case of linearly polarised GW(s) which in the two dimensional plane can be writen in terms of the Pauli matrices $\sigma^{1}$ and $\sigma^{3}$
\begin{equation}
h_{jk} \left(t\right) = 2f(t) \left(\varepsilon_{1}\sigma^1_{jk} 
+ \varepsilon_{3}\sigma^3_{jk}\right) 
\label{e13}
\end{equation}
where $\varepsilon_{1}$ and $\varepsilon_{3}$ represent the two possible time independent polarisation states of the GW satisfying the condition $\varepsilon_{1}^2+\varepsilon_{3}^2 = 1$. $2f(t)$ is the amplitude of the GW satisfying the boundary condition  
\begin{equation}
f(t)=0 \>, \quad {\rm for} \ t \le 0.
\label{bc}
\end{equation}
which physically signifies that the GW hits the particle at t=0. 

Now the matric-valued functions $\xi_{jk}$ and $\zeta_{jk}$, being $2\times 2$ complex matrices, can be expressed as linear combinations of the Pauli spin matrices and identity matrix  
\begin{eqnarray}
\zeta_{jk}\left(t \right) &=& A I_{jk} + B_{1}\sigma^{1}_{jk} + B_{2}\sigma^{2}_{jk} + B_{3}\sigma^{3}_{jk} 
\label{form2}\\
\xi_{jk} \left(t \right) &=&  C I_{jk} + D_{1} \sigma^{1}_{jk} + D_{2} \sigma^{2}_{jk} + D_{3} \sigma^{3}_{jk}
\label{form1}  
\end{eqnarray}
with $A, C, B_{a}, D_{a}, a = 1,2,3$, all complex quantities. Using these forms (\ref{form2}), (\ref{form1}) and (\ref{e13}) in eq.(s)(\ref{e21a}, \ref{e21b}) and comparing the coefficients of $I$ and $\sigma$-matrices from both sides, we obtain a set of first order differential equations for $A, B_{1}, B_{2}, B_{3}, C, D_{1}, D_{2}, D_{3}$ with appropriate boundary conditions consistant with (\ref{bc0}): 
\begin{eqnarray}
\dot{A} &=& i \Lambda B_{2}- \dot{f}\left(\varepsilon_{1}B_{1} 
+ \varepsilon_{3}B_{3}\right) -4i\frac{\Lambda}{\varpi} \dot{f}(\varepsilon_{3}D_{1}-\varepsilon_{1}D_{3})  \nonumber\\
\dot{B}_{1} &=& - \Lambda B_{3} - \dot{f}\left(\varepsilon_{1} A -
 i\varepsilon_{3}B_{2}\right) -4\frac{\Lambda}{\varpi} \dot{f}(i\varepsilon_{3}C-\varepsilon_{1}D_{2}) \nonumber\\
\dot{B}_{2}  &=& i \Lambda A - i \dot{f}\left(\varepsilon_{3}B_{1} - 
\varepsilon_{1}B_{3}\right)-4\frac{\Lambda}{\varpi} \dot{f}(\varepsilon_{1}D_{1}+\varepsilon_{3}D_{3})   \nonumber\\ 
\dot{B}_{3}&=& \Lambda B_{1}- \dot{f}\left(\varepsilon_{3} A 
+ i \varepsilon_{1}B_{2}\right)+4\frac{\Lambda}{\varpi} \dot{f}(i\varepsilon_{1}C+\varepsilon_{3}D_{2})   \nonumber\\ 
\dot{C} &=& - i \varpi A -i \Lambda D_{2}+ \dot{f}\left(\varepsilon_{1}D_{1} 
+ \varepsilon_{3}D_{3}\right)+4i\lambda \dot{f}(\varepsilon_{3}B_{1}-\varepsilon_{1}B_{3})  \nonumber \\ 
\dot{D}_{1} &=& - i \varpi B_{1}+ \Lambda D_{3}+ 
\dot{f}\left(\varepsilon_{1}C - i \varepsilon_{3}D_{2}\right) 
+4\lambda \dot{f}(i\varepsilon_{3}A-\varepsilon_{1}B_{2})\nonumber \\ 
\dot{D}_{2} &=& - i \varpi B_{2}- i \Lambda C + i \dot{f} \left(\varepsilon_{3}D_{1} 
- \varepsilon_{1}D_{3}\right)+4\lambda \dot{f}(\varepsilon_{1}B_{1}+\varepsilon_{3}B_{3})  \nonumber \\ 
\dot{D}_{3} &=& - i \varpi  B_{3}-  \Lambda D_{1} + 
\dot{f}\left(\varepsilon_{3} C + i \varepsilon_{1} D_{2}\right) 
-4\lambda \dot{f}(i\varepsilon_{1}A+\varepsilon_{3}B_{2})   
\label{iteration8} 
\end{eqnarray}
where 
\begin{eqnarray}
\Lambda = \frac{\bar{\theta}}{4m\hbar}
\label{Lambda}
\end{eqnarray}
is a frequency appearing naturally due to the presence of momentum noncommutativity in the system and 
\begin{eqnarray}
\lambda = \frac{m\varpi\theta}{4\hbar}
\label{lambda}
\end{eqnarray}
is a dimensionless parameter with spatial NC length-scale $\sqrt{\theta}$.

\noindent Solving the eq.(s)(\ref{iteration8}) to first order in the GW amplitude with boundary conditions (\ref{bc}), we get 
\begin{eqnarray}
A(t) &=& \cos\Lambda {t}\quad; \quad C(t) = -{i} \frac {\varpi}{\Lambda} \sin\Lambda{t} +  \cos\Lambda{t}\quad; 
\quad B_{2}(t) = - D_{2}(t) = i \sin\Lambda{t} \nonumber\\
B_{1}(t) &=& -\left\{ (\varepsilon_1 +4i\frac{\Lambda}{\varpi}\varepsilon_3)\cos\Lambda{t} + 5\varepsilon_3\sin\Lambda{t} \right\}f(t) + 5\Lambda \varepsilon_3 \int^{t}_{0} f(t')\cos\Lambda t'dt' \nonumber\\
B_{3}(t)  &=& \left\{ 5\varepsilon_1\sin\Lambda{t} - (\varepsilon_3 -4i\frac{\Lambda}{\varpi}\varepsilon_1)\cos\Lambda{t} \right\}f(t)  - 5\Lambda\varepsilon_1 \int^{t}_{0} f(t')\cos\Lambda t' dt' \nonumber\\
D_{1}(t) &=& \left\{ (\varepsilon_1 +4i\lambda\varepsilon_3)\cos\Lambda{t} - (\varepsilon_3 + i\frac{\varepsilon_1\varpi}{\Lambda})\sin\Lambda{t} \right\}f(t)+\left(\Lambda\varepsilon_3 + i\varepsilon_1\varpi\right) \int^{t}_{0} f(t')\cos\Lambda t' dt' \nonumber\\
D_{3}(t)  &=& \left\{ (\varepsilon_3-4i\lambda\varepsilon_1)\cos\Lambda{t}+(\varepsilon_1- i\frac{\varepsilon_3\varpi}{\Lambda})\sin\Lambda{t}\right\}f(t)
-(\Lambda\varepsilon_1 -i\varpi\varepsilon_3) \int^{t}_{0}  f(t')\cos\Lambda t' dt' .\nonumber\\   
\label{100z}
\end{eqnarray}
Using  the above expressions and eq(s)(\ref{form2}, \ref{form1}, \ref{e19}, \ref{xp}) yields the expectation value of the components of position and momentum of the particle at any arbitrary time $t$ in terms of the initial expectation values of the position $\left(X_{1}\left( 0 \right), X_{2}\left( 0 \right)\right)$ and momentum $\left(P_{1}\left( 0 \right), P_{2}\left( 0 \right)\right)$. We will give the explicit expression for $\langle X_{1}\left(t\right)\rangle$ and discuss the interesting features it presents.
\begin{eqnarray}
\langle X_{1}\left(t\right)\rangle & = & \left[ \cos\Lambda{t} +  \left(\varepsilon_3 \cos \Lambda t +\epsilon_1\sin\Lambda t\right)f(t) - \Lambda\varepsilon_1\int^{t}_{0} f(t')\cos\Lambda t' dt'\right]X_{1}\left( 0 \right)  \nonumber \\
 &+& \left[ \sin\Lambda{t} +  \left(\varepsilon_1\cos\Lambda t - \varepsilon_3 \sin \Lambda t\right)f(t)+ \Lambda\varepsilon_3\int^{t}_{0} f(t')\cos\Lambda t'dt' \right]  X_{2}\left( 0 \right) \nonumber\\ 
&+& \left[\frac{\varpi}{\Lambda}\sin\Lambda{t} + \left\{4\lambda\varepsilon_1 \cos\Lambda t + \varepsilon_3 \frac{\varpi}{\Lambda}\sin\Lambda{t}\right\}f(t) -\varpi\varepsilon_3\int^{t}_{0} f(t')\cos\Lambda t'dt' \right]\frac{P_{1}\left( 0 \right)}{m\varpi} \nonumber\\
&+& \left[\left\{\frac{\varpi}{\Lambda}\varepsilon_1 \sin\Lambda{t}-4\lambda\varepsilon_3 \cos\Lambda t \right\}f(t) -\varpi\varepsilon_1\int^{t}_{0} f(t')\cos\Lambda t'dt' \right]\frac{P_{2}\left( 0 \right)}{m\varpi}~.
\label{x1}
\end{eqnarray}
The above result reflect the effects due to both the NC parameters $\theta$ and $\bar\theta$ on the expectation
value of the coordinates of the particle. First let us note that the spatial noncommutativity leads to terms linear in the spatial NC parameters $\lambda$ and coupled to GW amplitude. Hence they will produce too small an effect to be detected at the present GW detector sensitivities as we have speculated earlier. Also, they will go undetected if GWs are not present. 

In contrast, momentum noncommutativity brings in an oscillatory nature in the dynamical evolution of the free particle with a frequency characterized by the momentum NC parameter $\Lambda = \frac{\bar{\theta}}{4m\hbar}$. Some of these oscillatory terms are independent of any GW interaction and hence will affect the particle's motion even when there are no GWs. This will lead to a noise source in the GW detectors which is not of any instrumental or geological/terrestrial origin but solely an effect of the phase-space NC algebra. Consequently, this noise will have same characteristics in all GW detector sites. Normally, the instrumental and geological/terrestrial noises in different GW detector read-outs are uncorrelated \cite{correlation} and this fact is used to eliminate such noise by cross-correlating data from two or more detectors at different sites. But any noise of NC momentum origin will stand out even after such elimination because of its global nature.

Also, the NC oscillatory noise may lead to a resonance behaviour if the detectors are indeed hit by a GW of the form $f\left(t\right) = h_{0} \sin \omega t$ with similar milli-hertz range frequency $\omega \approx \Lambda$. This is due to the presence of $\int^{t}_{0} f(t')\cos\Lambda t'dt'$ factors in (\ref{x1}). Such a resonance, if observed in all detector sites, may serve to pinpoint the characteristic NC frequency and therefore the value of the momentum NC parameter.

With the currently available upper-bound estimation of the momentun NC parameter \cite{samanta, bert0} $\bar{\theta} \leq 2.32 \times 10^{-61} {\rm kg^{2}m^{2}sec^{-2}}$ and taking neutron (mass $= 167.32 \times 10^{-29}$Kg) as the free particle the characteristic frequency turns out to be $\Lambda = 0.33 Hz$.  Interestingly the upcoming space-based GW detectors (e.g. LISA \cite{lisa}) will look precisely in the milli-Hz frequency range\footnote{So far the ground based detectors such as LIGO, Virgo, GEO600 and KAGRA are aimed to perform direct detection of GW in the 10-10000 Hz frequency band.}. Also a combined atom and laser interferometry technique has been proposed that will extend the gravitational wave detection frequency band to the lower-frequency region, namely 0.1-10Hz \cite{TOBA1, TOBA2}. Authors of this proposed new design argue that in this setting the known terrestrial source of  noises can be reduced such that measurements in this frequency band still can be accomplished using terrestrial detectors. Such detectors, when in place, will be  potential candidates for registering NC oscillatory noise with the characteristic frequency $\Lambda$. Though these detectors construction and design will be more complicated than a ``free particle's response to GW" scenario, the roll of momentum NC to introduce a characteristic frequency in these more realistic settings of GW detectors will still be plausible. Also, since theoretically modeling the atom interferometric experiments essentially involves quantum mechanics, our quantum mechanical treatment in the NC phase space will also be relevant in this connection.

To put the expression (\ref{x1}) in a form where various limiting situations can be easily read off we expand the oscillatory factors and retain terms linear in $\Lambda$. This gives 
\begin{eqnarray}
\langle X_{1}\left(t\right)\rangle & = &\left[X_{1}\left( 0 \right) + \frac{P_{1}\left( 0 \right)t}{m}\right]  + f\left(t\right) \left[\varepsilon_3\left\{ X_{1}\left( 0 \right) + \frac{P_{1}\left( 0 \right)t}{m}\right\} + \varepsilon_1 \left\{ X_{2}\left( 0 \right) + \frac{P_{2}\left( 0 \right)t}{m}\right\} \right]\nonumber\\
&& - \left[\frac{P_{1}\left( 0 \right)t}{m} \varepsilon_3 + \frac{P_{2}\left( 0 \right)t}{m} \varepsilon_1\right] \int^{t}_{0} f(t') dt'+ \left[ \frac{\theta P_{1}\left( 0 \right)}{\hbar} \varepsilon_1 - \frac{\theta P_{2}\left( 0 \right)}{\hbar} \varepsilon_3 \right] f\left(t\right) \nonumber\\ 
&& + \frac{\bar{\theta}}{4 m \hbar} \left[ t X_{2}\left( 0 \right) + \left\{\varepsilon_1  X_{1}\left( 0 \right) - \varepsilon_3  X_{2}\left( 0 \right)\right\} \left\{ t f\left(t\right) - \int^{t}_{0} f(t')  dt'\right\} \right]
\label{x1linear}
\end{eqnarray}
The first term shows the classical motion of the free particle with given initial position and momentum. In the absence of phase-space noncommutativity ($\theta = \bar{\theta} = 0$), and GW ($f\left(t\right) = 0$), this is the only surviving part. The second and third terms give the modulation caused by GW in the time evolution of the particle. In the limit $\theta \to 0$ and $\bar{\theta} \to 0$, these terms, along with the first term, give the usual response of the free particle to linearized GW. The fourth term shows the Bopp shift effect of spatial noncommutativity coupled to GW. Finally, the fifth term is the effect of momentum noncommutativity, part of which is coupled to the GW interaction, but a contribution independent of gravity also exists. This will result in a noise characterized by the momentum noncommutative parameter $\bar{\theta}$ and particle mass $m$. Since we have linearized in $\Lambda$ this shows the initial linearly increasing part of the $\sin \Lambda t$ curve with $t$. The identification of this noise in the GW data may also provide the evidence in favour of the existence of momentum noncommutativity and the parameter $\bar{\theta}$ can be estimated from the slope of the noise curve. 

\section*{Acknowledgemnet} 
The authors would like to thank the Reviewer for useful comments.
%%%%%%%%%%%%%

%%%%%%%%%%%%%%%%%%%%%%%%%%%%%%%%%%%%%%%%%%%%%%%%%%%%%%%%%%%%%%%%%%%%%%%%%%%%%%%%
\end{document}